\documentclass[twocolumn,secnumarabic,amssymb, nobibnotes, aps, prd]{revtex4-1}
\usepackage{graphicx}
\usepackage{subfigure}
\usepackage[T1]{fontenc}
\usepackage[utf8]{inputenc}
\usepackage[colorlinks,linkcolor=blue,anchorcolor=blue,citecolor=blue]{hyperref}
\usepackage{indentfirst}
\usepackage{amsmath}

\begin{document}

\title{Rescattering effects on spin-interference for $\rho^{0}$ photoproduction in heavy-ion collisions}
\author{Yusong Wang}

\author{Xinbai Li}
\author{Ziyang Li}
\author{Zebo Tang}
\email{zbtang@ustc.edu.cn}

\author{Xin Wu}

\author{Wangmei Zha}
\email{first@ustc.edu.cn}
\affiliation{State Key Laboratory of Particle Detection and Electronics, University of Science and Technology of China, Hefei 230026, China}

\begin{abstract}
\begin{@twocolumnfalse}

Recent measurements by various experiments in ultra-peripheral collisions have observed spin-interference in $\rho^{0}$ photoproduction, marking a breakthrough in Fermi-scale quantum interference experiments. Building on this, STAR extended the measurement to hadronic heavy-ion collisions, where significant rescattering effects on $\rho^{0}$ mesons were expected. In this study, we investigate how these rescattering effects influence the measurement of spin-interference. By embedding $\rho^{0}$ mesons produced via photoproduction, modeled by the Vector Meson Dominance model, into the Ultrarelativistic Quantum Molecular Dynamics framework, we estimate the impact on the $\cos2\phi$ and $\cos4\phi$ modulations, where $\phi$ is the angle between $\rho^{0}$ and one of the daughters' ($\pi^{\pm}$) transverse momentum. The results indicate a significant suppression of the $\cos2\phi$ modulation, while the $\cos4\phi$ modulation remains largely unaffected, which provides insight for understanding the difference due to rescattering effects between experimental measurements and theoretical predictions for $\rho^{0}$ photoproduction in heavy-ion collisions.

\end{@twocolumnfalse}
\end{abstract}

\maketitle

\section{Introduction}
\label{sec:1}

In relativistic heavy-ion collisions, the highly Lorentz-contracted electromagnetic fields surrounding the fast-moving nuclei serve as intense sources of quasi-real photons. A photon emitted from one nucleus can fluctuate into a quark-antiquark pair, which subsequently scatters off the other nucleus to emerge as a real vector meson~\cite{ref:1,ref:13.5}. This process, facilitated by elastic scattering via Pomeron exchange, confines the interaction to a specific site within one of the colliding nuclei~\cite{ref:2}, establishing a scenario analogous to a double-slit experiment. Furthermore, the photons from these Lorentz-contracted electromagnetic fields are expected to be fully linearly polarized, as suggested by Li et al.~\cite{ref:3} and later confirmed by the STAR Collaboration through dielectron measurements~\cite{ref:4}. This linear polarization can be inherited by the vector mesons produced via photoproduction, resulting in characteristic asymmetries in their decay angular distributions. Such asymmetries create an interference pattern in polarization space, offering a distinct signature of vector meson photoproduction in heavy-ion collisions. This spin-interference effect, originally proposed by Xing et al.~\cite{ref:5} and Zha et al.~\cite{ref:6,ref:7}, has since been experimentally validated by measurements from the STAR~\cite{ref:8,ref:9} and ALICE~\cite{ref:10,ref:11} collaborations, providing critical insights into quantum interference phenomena in these environments.

Traditionally, coherent photoproduction was thought to occur exclusively in ultra-peripheral collisions (UPCs), where the impact parameter \( b \) exceeds twice the nuclear radius, ensuring that the nuclei remain intact and meet the conditions required by coherent interactions. However, recent experimental observations have challenged this assumption. Measurements by the ALICE~\cite{ref:12} and STAR~\cite{ref:13} collaborations revealed an excess of J/$\psi$ production at very low transverse momentum (\( p_{\rm T} < 0.3 \) GeV/c) even in hadronic heavy-ion collisions (HHICs), where significant nuclear overlap occurs. This low-\( p_{\rm T} \) excess can be well-explained by theoretical models based on photoproduction processes, demonstrating that coherent photoproduction can indeed take place in HHICs. Beyond photon-nucleus interactions, ALICE~\cite{ref:14} and STAR~\cite{ref:15,ref:15.5} have also observed evidence of coherent photon-photon interactions in HHICs through measurements of dilepton production at similarly low \( p_{\rm T} \), which has been further verified by theoretical calculations incorporating photon-photon reactions~\cite{ref:2.5}. These findings greatly expand the scope for studying photon-induced reactions in relativistic heavy-ion collisions. They also offer novel approaches to probing the hot and dense medium formed in the nuclear overlap region.

Building on previous findings~\cite{ref:16}, STAR extended spin-interference measurements of $\rho$ photoproduction to that in hadronic heavy-ion collisions, aiming to explore the impact parameter dependence of the spin-interference effect in this more complex environment. Unlike the relatively clean conditions in UPCs, the overlapping region in hadronic collisions introduces intense strong interactions that may dilute or distort the underlying physical signals. Specifically, the $\rho$ meson, with a lifetime of approximately 1.3 fm/\(c\), typically decays before kinetic freeze-out, making it particularly sensitive to interactions with the dense hadronic medium. Experimentally, $\rho$ mesons are reconstructed via their decay into pion pairs. However, if a $\rho$ decays early, its decay products are likely to undergo rescattering with other hadrons, either through elastic or inelastic collisions. Such rescattering can modify the momentum of the decay daughters or absorb them entirely, thus preventing the experimental reconstruction of the parent resonance~\cite{ref:17,ref:18,ref:19}. Evidence for this significant rescattering effect has been observed in the centrality and transverse momentum (\(p_{\rm T}\)) dependence of \(K^{*0}\) production in heavy-ion collisions, where the \(K^{*0}\) meson, with an even longer lifetime than the $\rho$, undergoes similar rescattering effects~\cite{ref:20}.The probability of rescattering is closely related to the particle density profile, which in turn depends on the collision geometry~\cite{ref:21}. In non-central heavy-ion collisions, rescattering probabilities vary with azimuthal angle due to the initial collision geometry, potentially creating an anisotropic distribution. Since the spin-interference of $\rho$ mesons is evaluated through the angular distributions of their decay daughters relative to the $\rho$'s flight direction~\cite{ref:6}, which is strongly correlated with the reaction plane, rescattering effects could distort these distributions in reconstructable $\rho$ photoproduction events. Consequently, this effect could generate an artificial spin-interference signal, complicating the interpretation of the measurement.

This study examines the impact of rescattering effects on the measurement of spin-interference for $\rho^{0}$ photoproduction in HHICs. Using the Vector Meson Dominance (VMD) model to simulate photoproduced $\rho^{0}$ mesons, we embed these mesons within the Ultrarelativistic Quantum Molecular Dynamics (UrQMD) framework to evaluate the result for $\cos2\phi$ and $\cos4\phi$ angular modulations. By modeling this interaction within a realistic hadronic environment, we aim to capture how rescattering influences the observable spin-interference patterns in HHICs. Section~\ref{sec:2} details the methodology employed, while Sections~\ref{sec:3} and~\ref{sec:4} present the results and conclusions derived from our analysis.

\section{Methodology}
\label{sec:2}

To investigate the rescattering effects on photoproduced $\rho^{0}$ mesons in hadronic heavy-ion collisions, this study combines photoproduction modeling with the UrQMD framework. The methodology consists of three main components: the theoretical framework for photoproduction, the UrQMD model for simulating hadronic interactions in the overlap region, and the implementation of $\rho^{0}$ polarization to study angular modulations in decay products.

\subsection*{Theoretical Framework for Photoproduction}
\label{method-1}
In relativistic heavy-ion collisions, quasi-real photons generated by the intense Lorentz-contracted electromagnetic fields of colliding nuclei can fluctuate into virtual vector mesons, such as the $\rho^{0}$. These mesons subsequently scatter off nuclei and emerge as real particles. The photon flux density \( n(\omega_{\gamma}, \vec{x}_{\perp}) \) induced by the relativistic nuclei can be described by the Equivalent Photon Approximation (EPA)~\cite{ref:1}:

\begin{equation}
\label{eq:1}
\begin{split}
    n(\omega_{\gamma},\vec{x}_{\perp}) & =\frac{4Z^{2}\alpha}{\omega_\gamma}\Bigg|\int\frac{\,d^{2}\vec{k}_{\gamma\perp}}{(2\pi)^{2}}\vec{k}_{\gamma\perp}\frac{F_{\gamma}(\vec{k}_{\gamma})}{|\vec{k}_{\gamma}|^2}e^{i\vec{x}_{\perp}\cdot\vec{k}_{\gamma\perp}}\Bigg|^2,
\end{split}
\end{equation}
where \(\omega_{\gamma}\) is the photon energy, \( \vec{x}_{\perp} \) represents the transverse position from the nucleus center, and \(\vec{k}_{\gamma\perp}\) is the transverse momentum. Here, \( Z \) is the nuclear charge, \( \alpha \) the fine-structure constant, and \( F_{\gamma}(\vec{k}_{\gamma}) \) the electromagnetic form factor, derived from the Fourier transform of the nuclear charge density. The charge density follows the Woods-Saxon distribution:
\begin{equation}
\label{eq:2}
    \rho_{A}(r)=\frac{\rho_{0}}{1+\exp[(r-R_{WS})/d]},
\end{equation}
where \( R_{WS} = 6.38 \) fm is the nuclear radius for gold (Au), \( d = 0.535 \) fm is the skin depth~\cite{ref:22}, and \( \rho_{0} \) is the normalization factor. The photon momentum vector \( \vec{k}_{\gamma} \) can be expressed as:

\begin{equation}
\label{eq:3}
    \vec{k}_{\gamma}=\bigg(\vec{k}_{\gamma\perp},\frac{\omega_{\gamma}}{\gamma_{c}}\bigg),\quad\omega_{\gamma}=\frac{1}{2}\,M_{\rho}e^{\pm\,y},
\end{equation}
where \( \gamma_{c} \) is the Lorentz factor of the beam, and \( M_{\rho} \) and \( y \) are the mass and rapidity of the $\rho^{0}$ meson, respectively.

The scattering amplitude for \( \gamma A \rightarrow \rho^{0} A \) can be estimated by combining VMD approach~\cite{ref:23} with the Glauber model~\cite{ref:24}:

\begin{equation}
\label{eq:4}
    \Gamma(\gamma A \rightarrow \rho^{0} A) = \frac{f_{\gamma p \rightarrow \rho p}(0)}{\sigma_{\rho p}} 2 \bigg[1 - e^{-\frac{1}{2} \sigma_{\rho p} T'(\vec{x}_{\perp})}\bigg].
\end{equation}
Here \( f_{\gamma p \rightarrow \rho p}(0) \) is the forward-scattering amplitude:

\begin{equation}
\label{eq:5}
    f_{\gamma p \rightarrow \rho p}(0) = \sqrt{\frac{\,d\sigma(\gamma p \rightarrow \rho p)}{dt}\Big|_{t=0}}.
\end{equation}
and it can well parametrized from Ref.~\cite{ref:25}. $\sigma_{\rho p}$ is the total $\rho p$ cross section and can be extracted via VMD approach:

\begin{equation}
\label{eq.6}
    \sigma_{\rho p}=\frac{f_{\rho}}{4\sqrt{\alpha}C}f_{\gamma p\rightarrow\rho p}
\end{equation}
where the $\rho-\gamma$ coupling $f_{V}$ and the correction factor $C$ for the off-diagonal diffractive interaction can be obtained from Ref.~\cite{ref:26}. Additionally, the modified thickness function \( T'(\vec{x}_{\perp}) \) accounts for coherence length effects:

\begin{equation}
\label{eq:7}
    T'(\vec{x}_{\perp}) = \int \rho \big(\sqrt{|\vec{x}_{\perp}|^{2} + z^{2}}\big) e^{iq_{L}z} \, dz, \quad q_{L} = \frac{1}{2} \frac{M_{\rho}}{\gamma_{c}} e^{y},
\end{equation}
where \( q_{L} \) is the longitudinal momentum transfer necessary to produce a real $\rho^{0}$ meson. 

For $\rho^{0}$ production, we have to consider the resonant cross section for $\rho^{0}$ because of its larger width. In this research, it can be described by a Breit-Wigner resonant amplitude:

\begin{equation}
\label{eq:8}
    f_{\rho^{0}\rightarrow\pi^{+}\pi^{-}}=\frac{f_{0}\sqrt{\Gamma_{\rho}M_{\rho}M_{\pi\pi}}}{M_{\pi\pi}^{2}-M_{\rho}^{2}+i\Gamma_{\rho}M_{\rho}},
\end{equation}
where $f_{0}$ can be obtained from Ref.~\cite{ref:27}, and the Breit-Weigner width $\Gamma_{\rho}=\Gamma_{0}(p_{\pi}/p_{0})^{3}(M_{\rho}/M_{\pi\pi})$, where $p_{\pi}$ is the dacay daughters momentum in the $\rho^{0}$ rest frame, $p_{0}=358$ GeV/c is the $\pi$ momentum for $M_{\pi\pi}=M_{\rho}$ and $\Gamma_{0}$ is the pole $\rho^{0}$ width. We need cut off the range of $2m_{\pi}<M_{\pi\pi}<M_{\rho}+5\Gamma_{0}$ because it falls rapidly at large $M_{\pi\pi}$ according to most studies of $\gamma p\rightarrow\rho p$. 

The transverse spatial amplitude distribution for photoproduction is then expressed as~\cite{ref:28}:

\begin{equation}
\label{eq:9}
\begin{split}
    A(\vec{x}_{\perp}) = f_{\rho^{0}\rightarrow\pi^{+}\pi^{-}} \cdot \sqrt{n(\omega_{\gamma}, \vec{x}_{\perp}) \cdot \frac{\,d^{2}\sigma(\gamma A \rightarrow \rho^{0} A)}{\,d^{2} \vec{x}_{\perp}}}.
\end{split}
\end{equation}
where $f_{\rho^{0}\rightarrow\pi^{+}\pi^{-}}$ needs be normalized based on Eq.~\ref{eq:8}, and the spatial distribution can be expressed as:

\begin{equation}
\label{eq:10}
    \frac{\,d^{2}P}{\,d^{2}\vec{x}_{\perp}} = \Big|A_{1}(\vec{x}_{\perp}) + A_{2}(\vec{x}_{\perp}) \Big|^{2},
\end{equation}
Meanwhile, the corresponding momentum distribution is obtained by Fourier transformation to the spatial amplitude distribution:

\begin{equation}
\label{eq:11}
    \frac{\,d^{2}P}{\,dp_{x} dp_{y}} = \frac{1}{2\pi} \Bigg|\int d^{2}\vec{x}_{\perp} \, \big[A_{1}(\vec{x}_{\perp}) + A_{2}(\vec{x}_{\perp})\big] e^{i \vec{p}_{\perp} \cdot \vec{x}_{\perp}}\Bigg|^{2},
\end{equation}
where \( A_{1}(\vec{x}_{\perp}) \) and \( A_{2}(\vec{x}_{\perp}) \) denote spatial amplitude distributions for the two colliding nuclei, and these definitions can ensure uncertainty principle. Thus, both the transverse position and momentum information of $\rho^{0}$ mesons produced through photoproduction can be simulated according to these distributions.

\subsection*{UrQMD Framework for Modeling Hadronic Interactions}
\label{UrQMD_Intro}
The rescattering of photoproduced $\rho^{0}$ mesons in the dense hadronic medium created from the overlap region is simulated through the UrQMD model, a microscopic transport model based on string fragmentation and hadronic scattering processes. Each hadronic scattering cross section is determined by the principle of detailed balance~\cite{ref:29,Ref:30}. In the output files (f15 files), UrQMD provides a detailed record of each scattering event, including the type and cross section of each scattering, as well as some particles' information of the initial and final states, such as position, momentum, particle type, etc. This enables a comprehensive analysis of rescattering effects, allowing for the identification of reconstructable and scattered $\rho^{0}$ mesons from collision events, which will be clearly defined in the following paragraph. 

Typically, $\rho^{0}$ mesons are reconstructed through the decay channel $\rho^{0} \rightarrow \pi^{+} + \pi^{-}$, which has a branch ratio of nearly 100\%. The decay products, $\pi^{+}$ and $\pi^{-}$, may undergo additional elastic or inelastic scattering, resulting in the definition of three classifications of $\rho^{0}$ mesons:

\begin{enumerate}
    \item \textbf{Reconstructable $\rho^{0}$}: $\pi^{+}$ and $\pi^{-}$ from this decay channel remain unscattered throughout the evolution of the hadron gas.
    \item \textbf{Scattered $\rho^{0}$}: $\pi^{+}$ and $\pi^{-}$ from this decay channel experience scattering with other hadrons during the evolution of the hadron gas. In this case, the daughter particles will undergo rescattering.
    \item \textbf{All $\rho^{0}$}: All events in this decay channel, regardless of whether $\pi^{+}$ and $\pi^{-}$ undergo scattering, i.e., both reconstructable $\rho^{0}$ and scattered $\rho^{0}$.
\end{enumerate}

Since the UrQMD model only considers hadronically produced $\rho^{0}$ mesons, modifications to the initialization process are required to include externally generated photoproduced $\rho^{0}$ mesons. These external inputs must include parameters such as position coordinates, embedding time, momentum, energy, mass, particle ID, and charge information for each $\rho^{0}$ meson. The differential distributions for photoproduced $\rho^{0}$ mesons can be calculated by the framework outlined in the preceding subsection (Eq.~\ref{eq:10},~\ref{eq:11}). Each $\rho^{0}$ is assigned a unique ID number to differentiate between distinct events in subsequent reconstructions.

In each generated event, the collision times for the next interaction can be calculated using kinematic methods, assuming the particles move in uniform rectilinear motion. All possible collision modes (including decays) are considered, and then these events are ordered chronologically. The event with the shortest collision time is selected as the actual collision, and dynamic properties such as the cross section or decay width are calculated accordingly. The collision time from the actual event is then used to initialize the subsequent evolution, continuing this process iteratively until the final evolution stage is complete. For embedded $\rho^{0}$ mesons, controlling the collision timing is critical due to the presence of the embedding time (i.e., the time at which photoproduced $\rho^{0}$ mesons begin to participate in hadronic interactions within the nuclear collision background). Within the UrQMD code, evolution time adjustments can specify whether each $\rho^{0}$ meson should undergo collision in each simulation iteration, ensuring accurate modeling of the photoproduction and hadronic scattering processes.

\subsection*{Implementation of $\rho^{0}$ Polarization and Angular Modulation Analysis}
\label{method-3}

Photons generated from the highly Lorentz-contracted electromagnetic fields in relativistic heavy-ion collisions are expected to exhibit full linear polarization. This polarization is transferred to the photoproduced $\rho^{0}$ mesons, which, together with the double-slit interference effect arising from the two colliding nuclei, results in distinct asymmetries in the angular distributions of the decay products~\cite{ref:6}. These angular distributions can be systematically studied in a well-defined reference frame.
In this frame, the direction of the $\rho^{0}$ momentum $\vec{p}$ in the photon-nucleon center-of-mass frame is chosen as the $\hat{z}$ axis. The $\hat{y}$ axis is defined as normal to the photoproduction plane ($\hat{y} = \hat{k} \times \hat{z}$), where $\hat{k}$ represents the direction of the quasi-real photon involved in the photoproduction process. The $\hat{x}$ axis, serving as the quantization axis, is determined by $\hat{x} = \hat{y} \times \hat{z}$. The decay polar angle $\theta$ and azimuthal angle $\phi$ are then defined as follows~\cite{ref:31}:

\begin{equation}
    \label{eq:21}
    \cos\theta = \hat{\pi} \cdot \hat{z},
\end{equation}

\begin{equation}
    \label{eq:22}
    \cos\phi = \frac{\hat{y} \cdot (\hat{z} \times \hat{\pi})}{|\hat{z} \times \hat{\pi}|}, \quad \sin\phi = -\frac{\hat{x} \cdot (\hat{z} \times \hat{\pi})}{|\hat{z} \times \hat{\pi}|},
\end{equation}
where $\hat{\pi}$ represents the direction of one of the decay daughters, $\pi^{\pm}$, in the $\rho^{0}$ rest frame. In experimental measurements, the azimuthal angle $\phi$ is often approximated as~\cite{ref:16}:

\begin{equation}
    \label{eq:23}
    \cos\phi = \frac{(\vec{p}_{T1} + \vec{p}_{T2}) \cdot (\vec{p}_{T1} - \vec{p}_{T2})}{|\vec{p}_{T1} + \vec{p}_{T2}| \cdot |\vec{p}_{T1} - \vec{p}_{T2}|},
\end{equation}
where $\vec{p}_{T1}$ and $\vec{p}_{T2}$ are the transverse momenta of the decay daughters, $\pi^{\pm}$, in the laboratory frame. For all $\rho^{0}$, since the final state $\pi^{\pm}$ is scattered, we cannot directly extract the $\cos 2\phi$ and $\cos 4\phi$ modulation from the final state $\pi^{\pm}$. Consequently, the magnitude of $\cos 2\phi$ is determined based on the kinematic conditions at the beginning of the decay event (a moment before the final state $\pi^{\pm}$ scattered). Furthermore, since the reconstructable $\rho^{0}$ remains unaffected, the magnitude of $\cos 2\phi$ can still be calculated using the same definition as all $\rho^{0}$ as before.

The linear polarization of the photoproduced $\rho^{0}$ mesons manifests in the azimuthal angle distribution of the decay products ~\cite{ref:32}:

\begin{equation}
    \label{eq:24}
    \frac{\,dN}{\,d\phi} = \frac{1}{2\pi}\,(1 + P_{\gamma} \cos2\phi),
\end{equation}
where $P_{\gamma}$ represents the degree of polarization along the transverse momentum of the $\rho^{0}$. The degree of linear polarization can be extracted by:

\begin{equation}
    \label{eq:25}
    P_{\gamma} = \Bigg\langle\frac{A_{x}^{2} - A_{y}^{2}}{A_{x}^{2} + A_{y}^{2}}\Bigg\rangle,
\end{equation}
where $A_{x}$ and $A_{y}$ are the amplitudes along and perpendicular to the transverse momentum of the $\rho^{0}$, respectively. These amplitudes can be calculated by the theoretical framework detailed in subsection~\ref{method-1}.

For photoproduced $\rho^{0}$ mesons embedded within the UrQMD framework, decays are controlled by the default UrQMD process, which does not include decay anisotropy. However, modulations such as $\cos2\phi$ and $\cos4\phi$ can be readily implemented by applying appropriate weights to the decay angles during the analysis of the final decay products. This approach enables the study of angular distributions reflecting the intrinsic linear polarization of photoproduced $\rho^{0}$ mesons. To align with the simulation procedure in UrQMD, the decay angles can be weighted in the following way: First, select some events of $\rho^{0}$ decay from all $\rho^{0}$ obtained from photoproduction for the case of random polarization. Using the $\cos2\phi$ modulation as the function of $p_{T}$ obtained by the method in~\cite{ref:32}, we can further select these events using Monte Carlo simulation and ensure the decay angles from them follow the distribution given by Eq.~\ref{eq:24}. After this selecting, identify the events that correspond to the reconstructable $\rho^{0}$ or not. The selected events from reconstructable $\rho^{0}$ can then be used to obtain the weighted distribution for reconstructable $\rho^{0}$.

\section{Result}
\label{sec:3}

\begin{figure*}
    \centering
    \begin{minipage}{1.0\textwidth}
    \centering
    \includegraphics[width=0.45\textwidth]{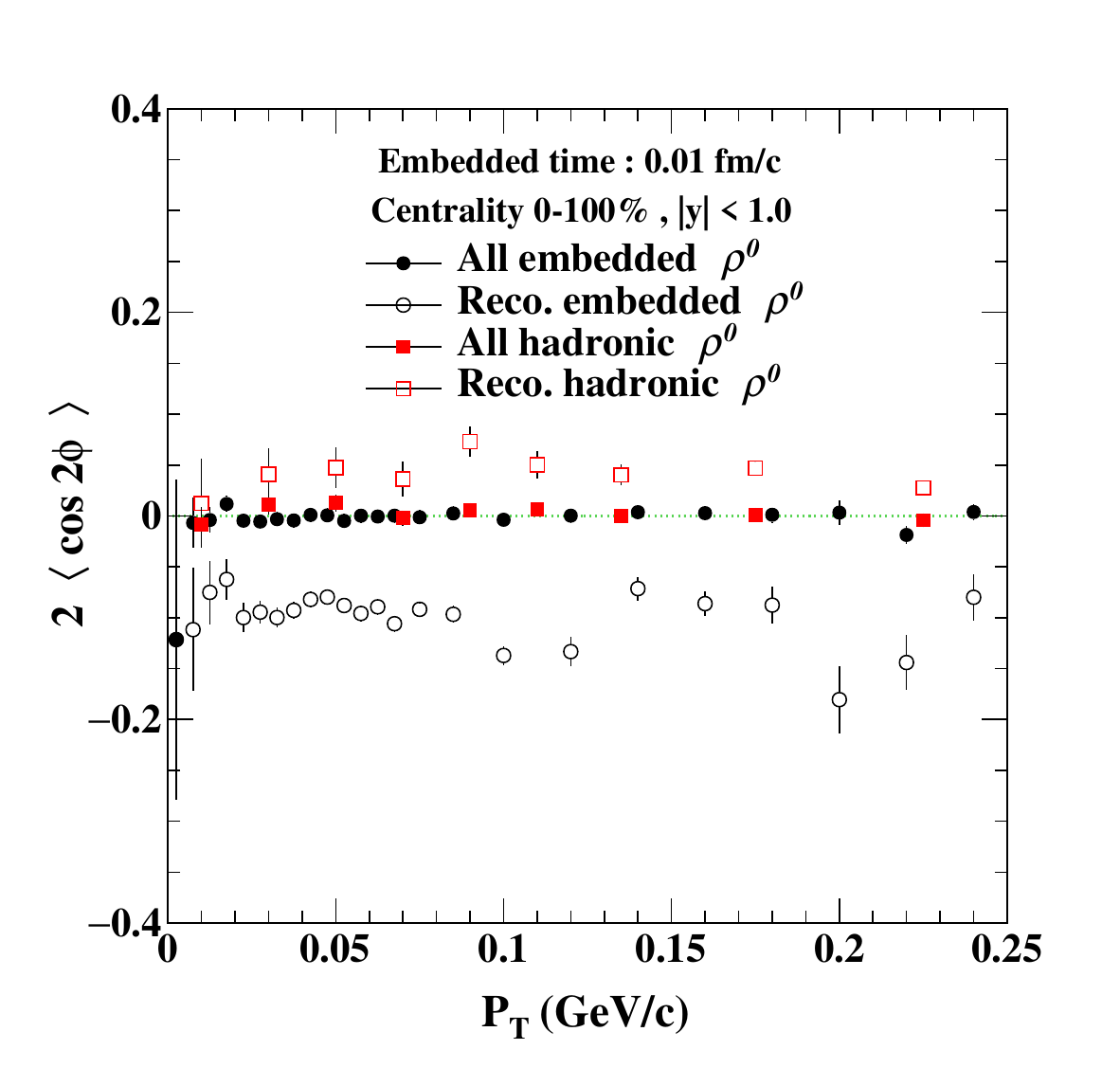}
    \centering
    \includegraphics[width=0.45\textwidth]{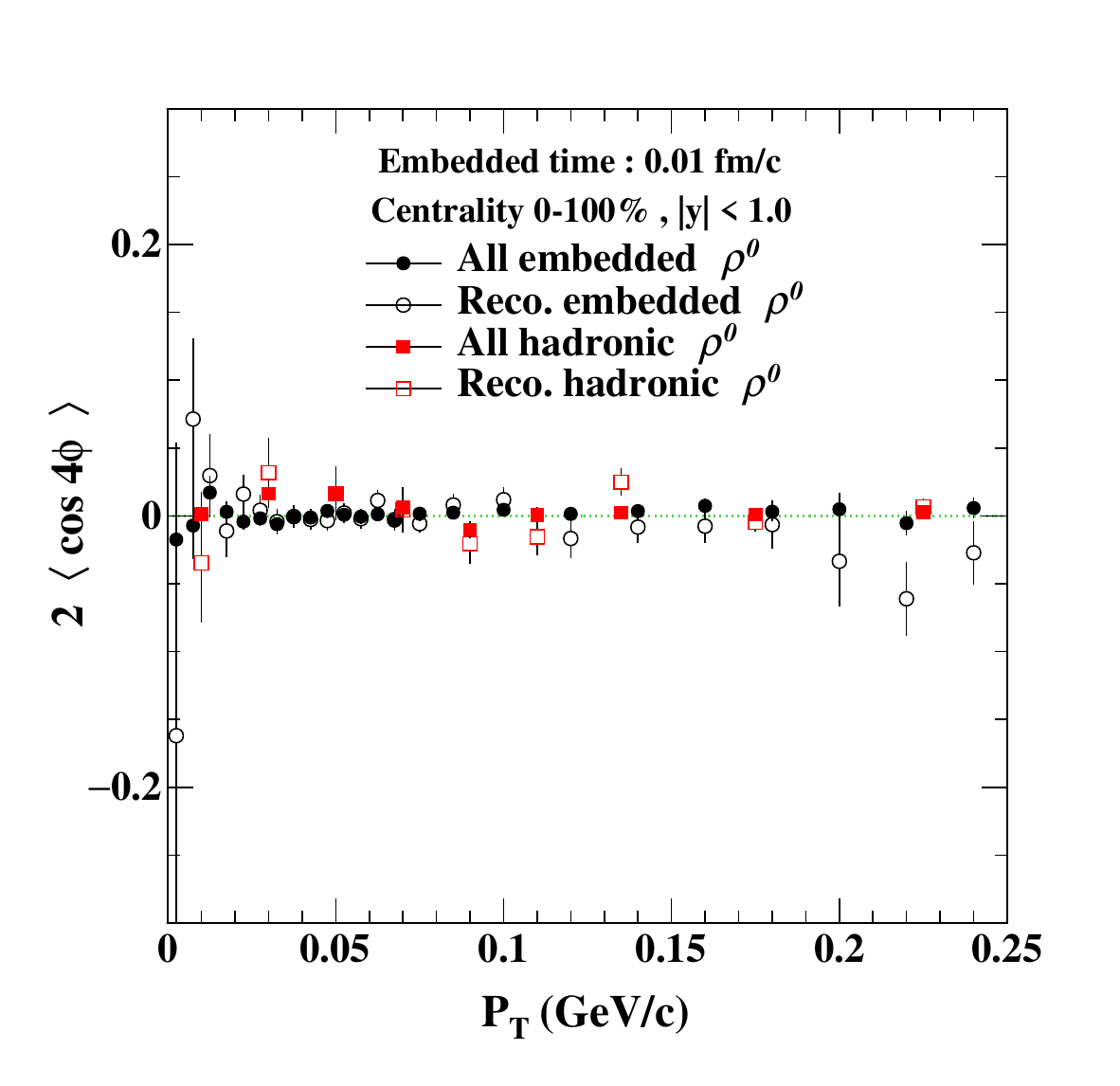}
    \end{minipage}
    \caption{The $\cos 2\phi$ and $\cos 4\phi$ modulations as a function of $p_{T}$ for both reconstructable and all, hadronic and photoproduced $\rho^{0}$ mesons at $0$–$100\%$ centrality in $\mathrm{Au}+\mathrm{Au}$ collisions at $\sqrt{s_{NN}} = 200$ GeV. The simulation is conducted  for the case of random polarization for $\rho^{0}$ at midrapidity ($|y| <1.0$) with an embedding time of 0.01 fm/c.}
    \label{fig:1}
\end{figure*}

Before implementing modifications to incorporate the rescattering effect for $\rho^{0}$ photoproduction, we applied the original UrQMD model to simulate the rescattering effect for hadronic $\rho^{0}$ mesons.  Fig.~\ref{fig:1} shows the $\cos 2\phi$ and $\cos 4\phi$ modulations for all hadronic $\rho^{0}$ mesons ("all $\rho^{0}$") and those that can be reconstructed ("reconstructable $\rho^{0}$"), as a function of transverse momentum ($p_{T}$) at $0-100\%$ centralities. The definitions of "all $\rho^{0}$" and "reconstructable $\rho^{0}$" are provided in Section~\ref{UrQMD_Intro}. The results indicate that the $\cos 2\phi$ and $\cos 4\phi$ modulations for all hadronic $\rho^{0}$ are zero, suggesting that the UrQMD model does not incorporate polarization effects during $\rho^{0}$ decay. However, for reconstructable $\rho^{0}$, a positive $\cos 2\phi$ modulation is observed, which implies that the rescattering effects on all $\rho^{0}$ indeed influence the $\cos 2\phi$ modulation, and is manifested as the residual effect in the case of reconstructable $\rho^{0}$. This effect is likely driven by the influence of asymmetric medium on pions, leading to a non-trivial modification of the decay angular distribution of $\rho^{0}$. In contrast, the $\cos 4\phi$ modulation for reconstructable $\rho^{0}$ remains zero, indicating that the rescattering effect on all $\rho^{0}$ may have little to no impact on $\cos 4\phi$ modulation. This suggests that the mechanism responsible for the anisotropic $\cos 2\phi$ distribution does not significantly contribute to higher order modulations, possibly due to the nature of the medium's asymmetry or the limited sensitivity of $\cos 4\phi$ to pions rescattering.

\begin{figure*}
    \centering
    \begin{minipage}{1.0\textwidth}
    \centering
    \includegraphics[width=0.45\textwidth]{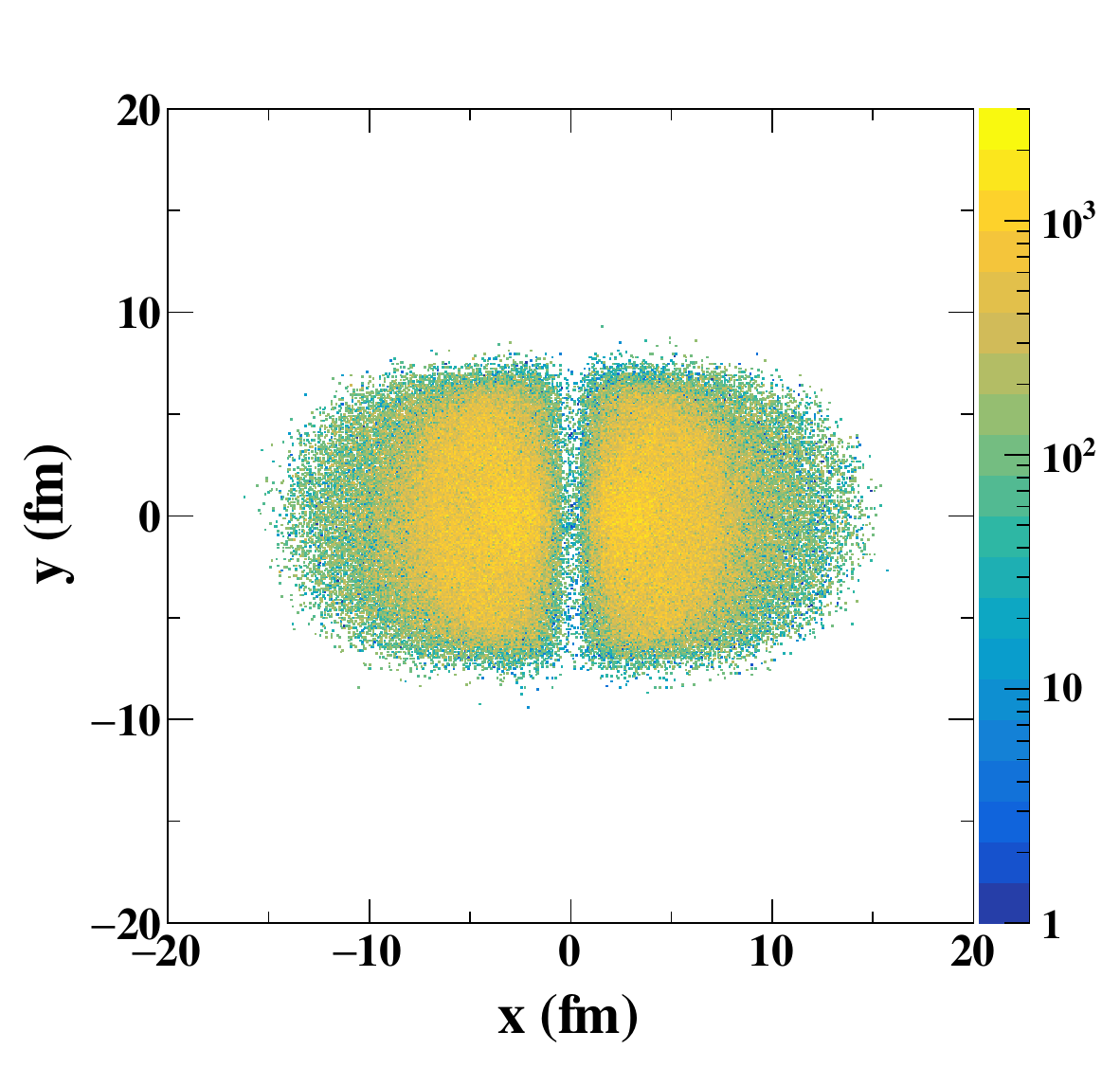}
    \centering
    \includegraphics[width=0.45\textwidth]{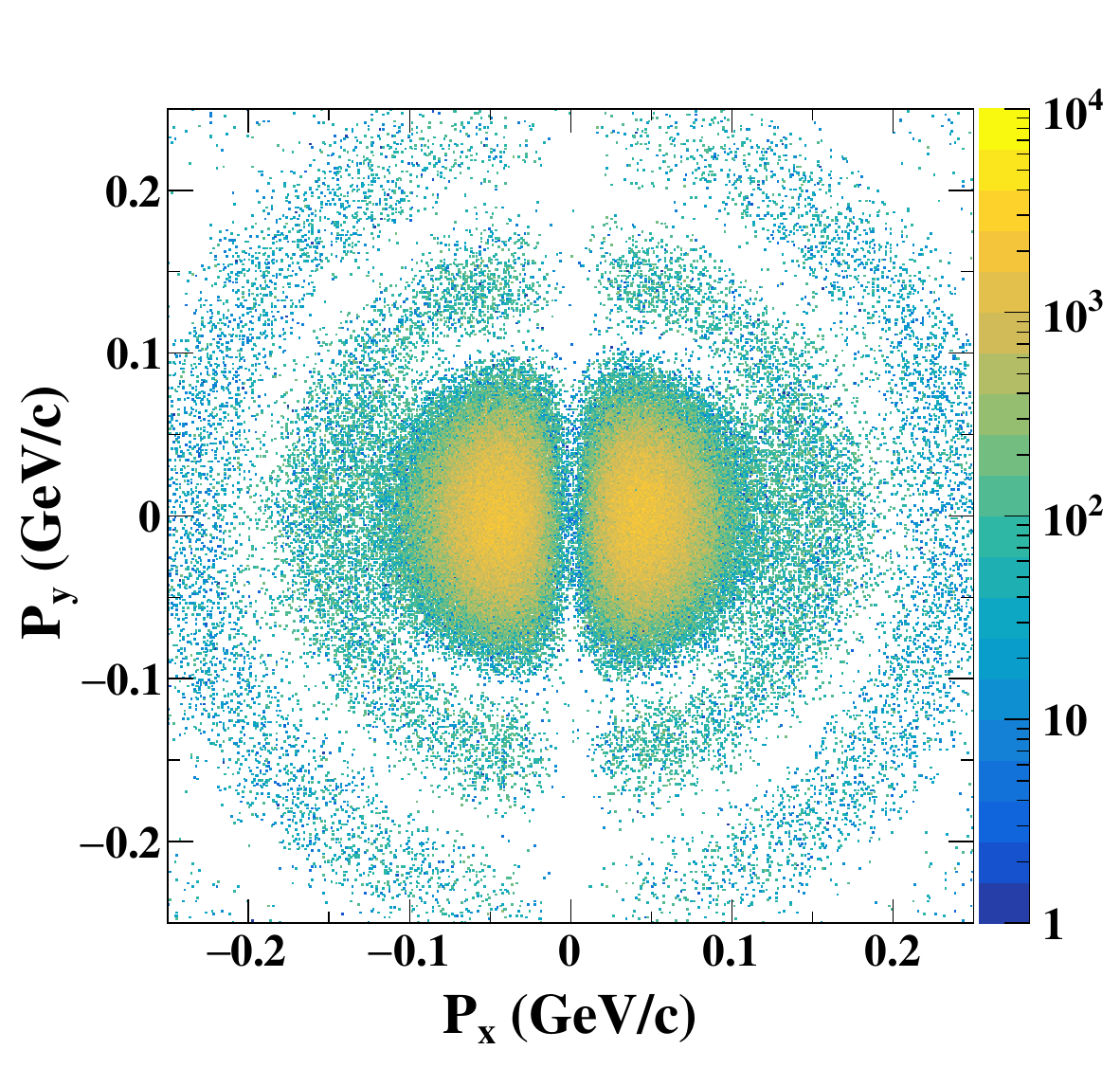}
    \end{minipage}
    \caption{Left Panel: The spatial distribution of embedded $\rho^{0}$ mesons in the plane perpendicular to beam direction ($x$-$y$ plane). Right Panel: The transverse momentum ($p_{T}$) distribution of embedded $\rho^{0}$ mesons in the $x$-$y$ plane.}
    \label{fig:2}
\end{figure*}

\begin{figure*}
    \centering
    \begin{minipage}{1.0\textwidth}
    \centering
    \includegraphics[width=0.45\textwidth]{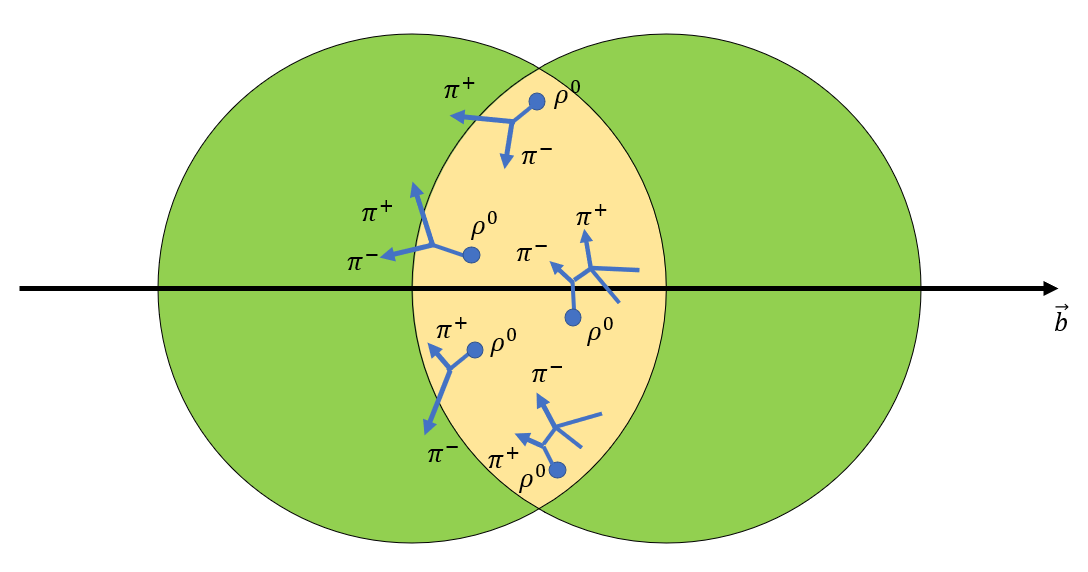}
    \centering
    \includegraphics[width=0.45\textwidth]{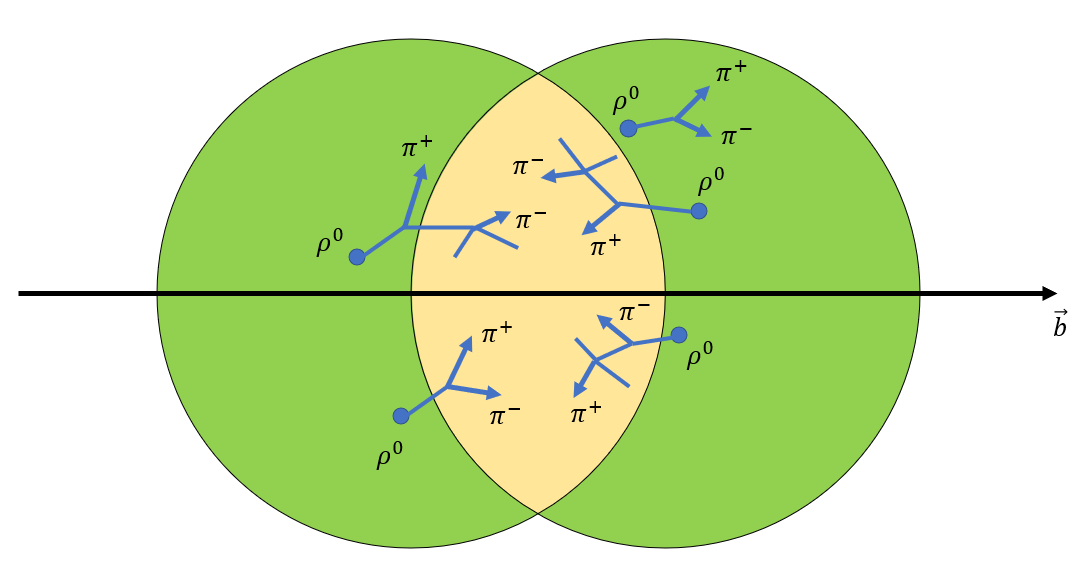}
    \end{minipage}
    \caption{A simplified illustration for the rescattering effect of the decay daughters from hadronic $\rho^{0}$ (left panel) and photoproduced $\rho^{0}$ (right panel). The big circles represent a pair of $\mathrm{Au}$ nuclei.}
    \label{fig:3}
\end{figure*}

\begin{figure*}
    \centering
    \begin{minipage}{1.0\textwidth}
    \centering
    \includegraphics[width=0.45\textwidth]{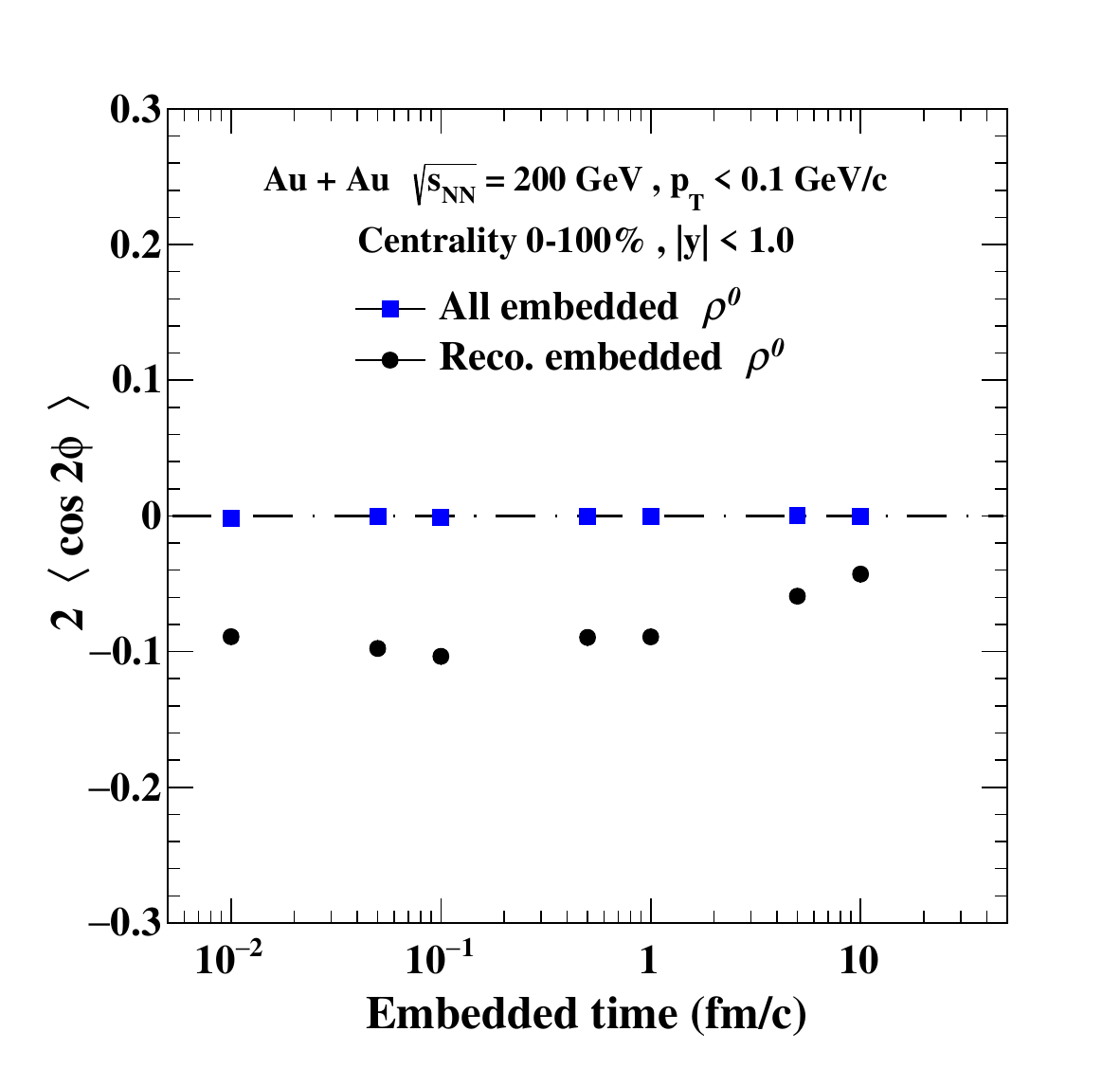}
    \centering
    \includegraphics[width=0.45\textwidth]{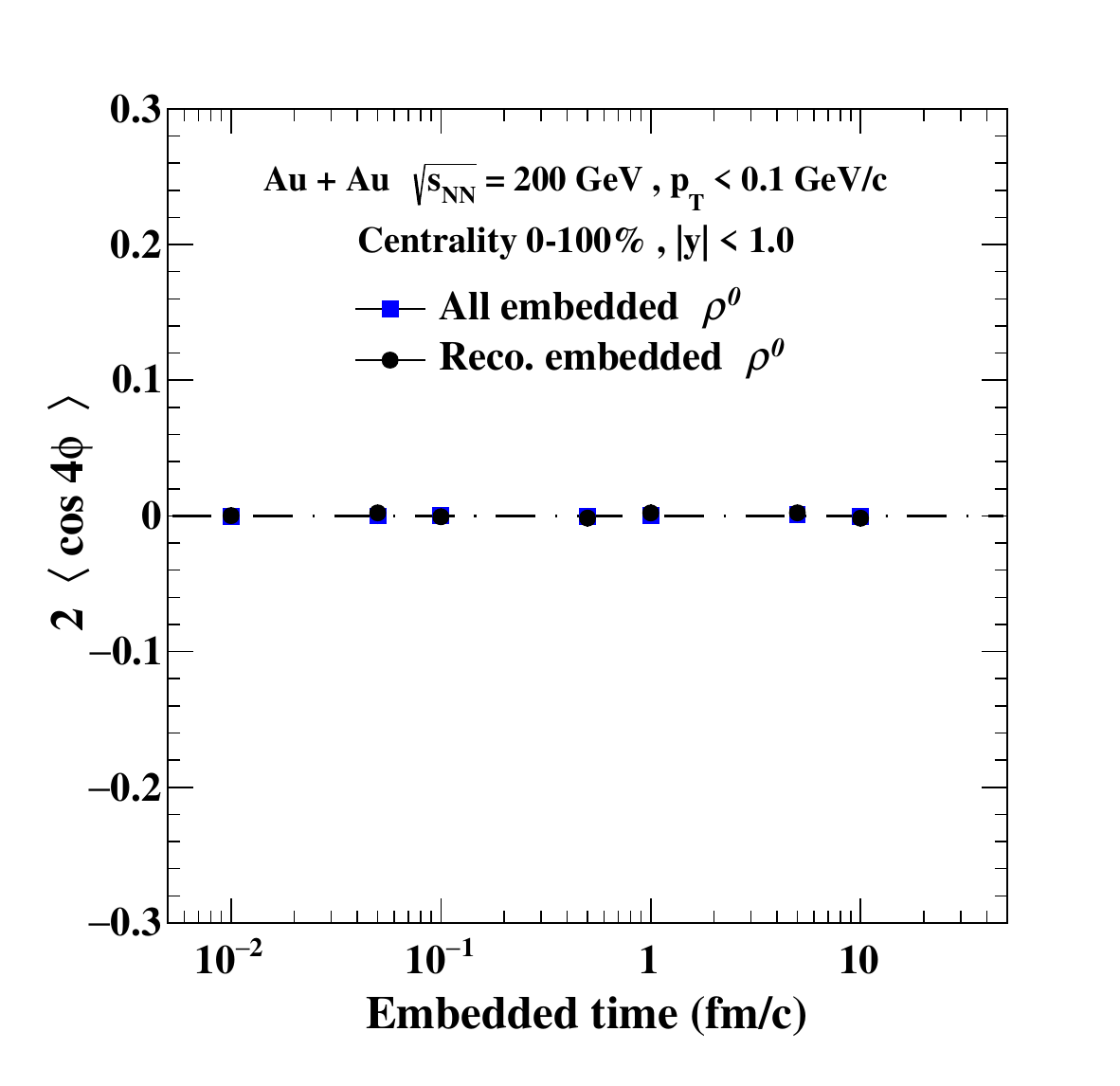}
    \end{minipage}
    \caption{The $\cos 2\phi$ and $\cos 4\phi$ modulations as functions of embedding time for reconstructable and all photoproduced $\rho^{0}$ mesons at $0$–$100\%$ centrality in $\mathrm{Au}+\mathrm{Au}$ collisions at $\sqrt{s_{NN}} = 200$ GeV at midrapidity for $p_{T} < 0.1$ GeV/c, under the assumption of random polarization.}
    \label{fig:4}
\end{figure*}
\begin{figure*}
    \centering
    \begin{minipage}{1.0\textwidth}
    \centering
    \includegraphics[width=0.45\textwidth]{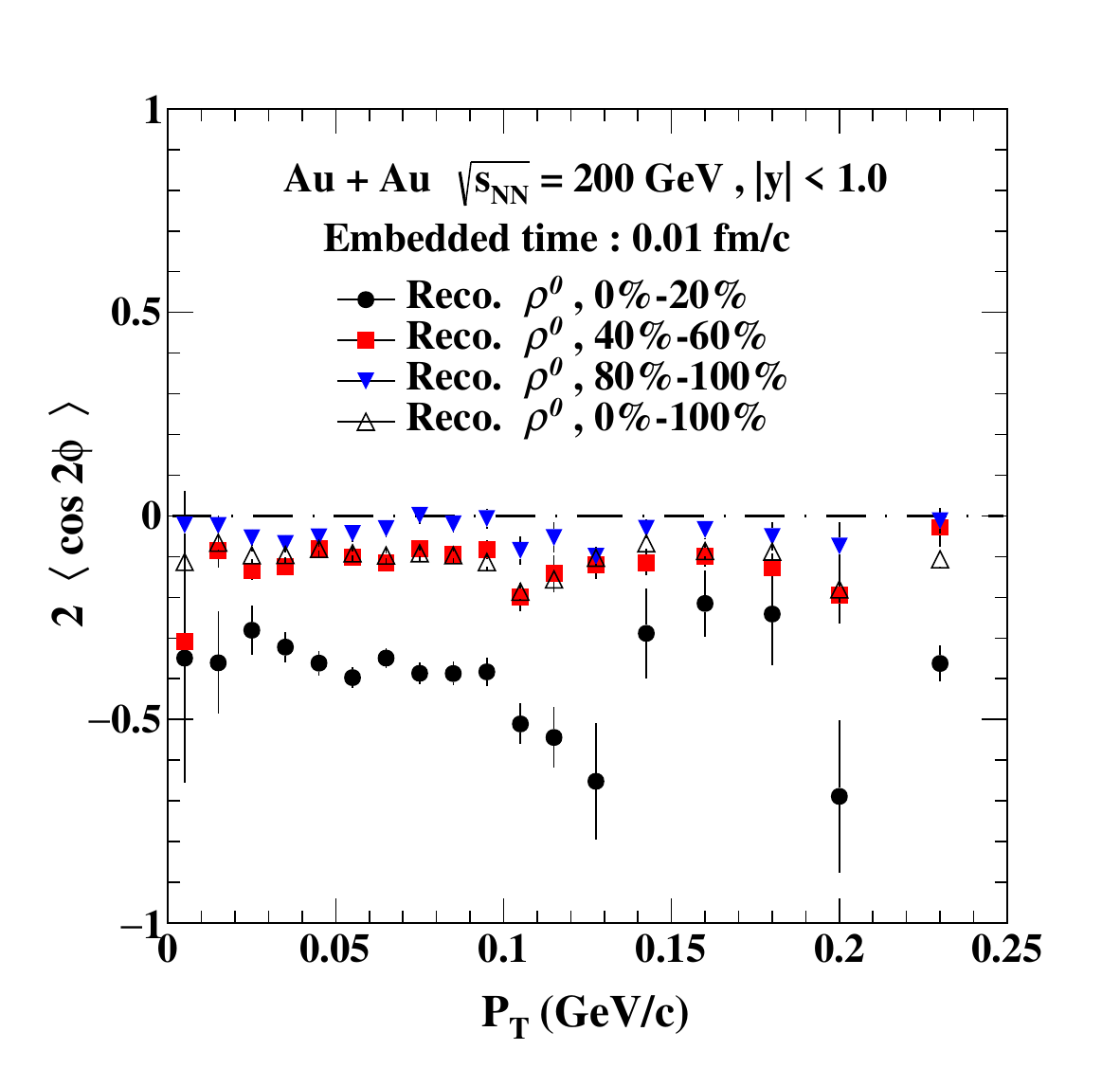}
    \centering
    \includegraphics[width=0.45\textwidth]{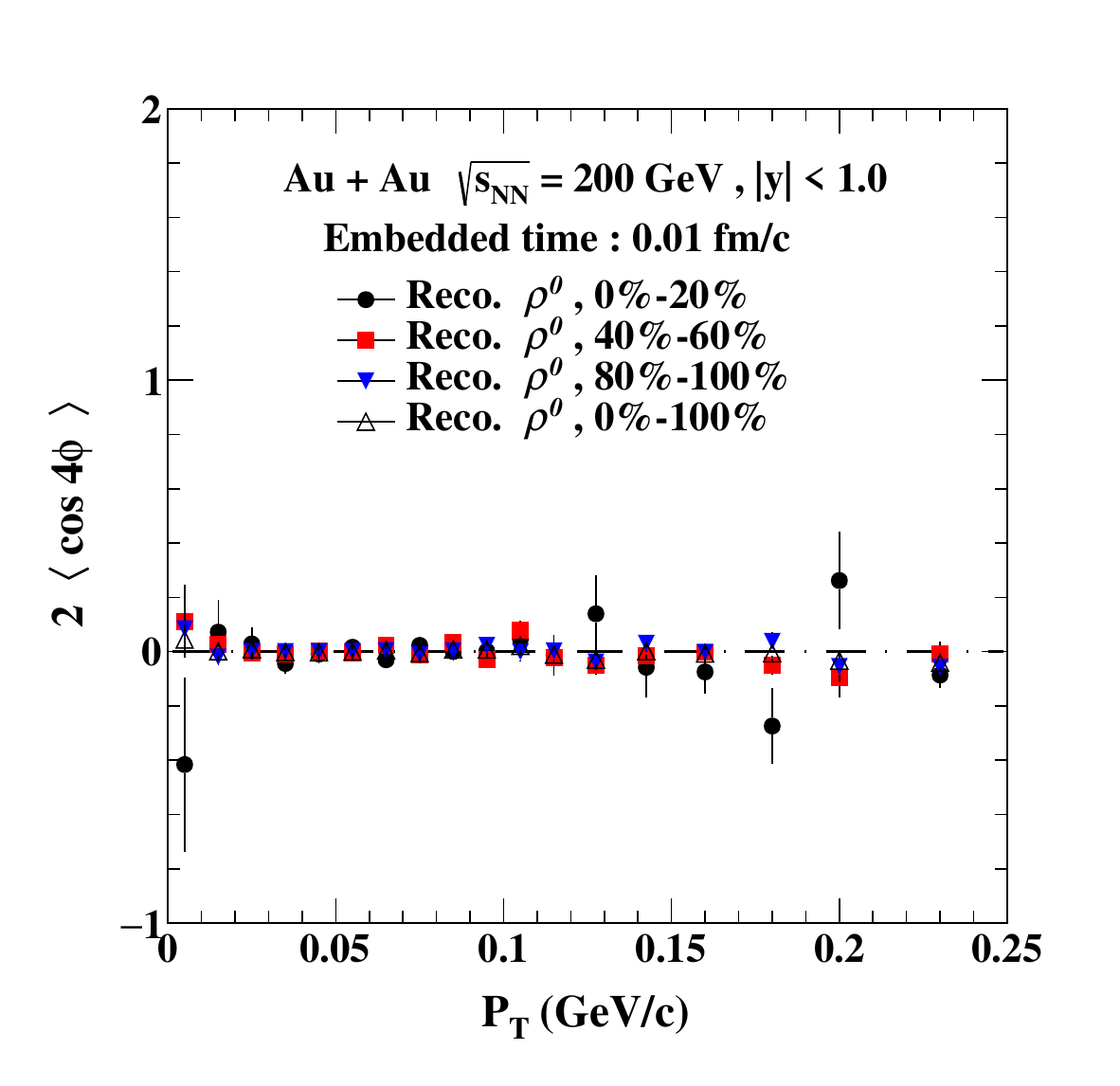}
    \end{minipage}
    \caption{The $\cos2\phi$ and $\cos4\phi$ modulation for the case of random polarization as functions of $p_{T}$ for reconstructable photoproduced $\rho^{0}$ mesons at $0-20\%$ (central), $40\%-60\%$ (semi-central), $80\%-100\%$ (peripheral), and $0-100\%$ centralities in $\mathrm{Au}+\mathrm{Au}$ collisions at $\sqrt{s_{NN}}=200$ GeV at midrapidity with the fixed embedded time 0.01 fm/c.}
    \label{fig:5}
\end{figure*}

To investigate the rescattering effect on photoproduced $\rho^{0}$ mesons, we modified the original UrQMD code to incorporate initial $\rho^{0}$ distributions generated through photoproduction, following the methodology outlined in Section~\ref{sec:2}. The $\cos 2\phi$ and $\cos 4\phi$ modulations for photoproduction with random polarization (no spin-int) as a function of $p_{T}$ at an embedding time of 0.01 fm/c are presented in Fig.~\ref{fig:1}. The decay of embedded $\rho^{0}$ is governed by the UrQMD program, corresponding to the case of production with random polarization. Consequently, the $\cos 2\phi$ and $\cos 4\phi$ modulations for all embedded $\rho^{0}$ are both zero. For reconstructable $\rho^{0}$, the results differ significantly between photoproduction and hadronic scenarios. Photoproduction exhibits a strong negative $\cos 2\phi$ modulation, whereas hadronic production produces positive values. Meanwhile, in both cases, the $\cos 4\phi$ modulation remains absent. As $p_{T}$ increases, two distinct dips emerge in the $\cos 2\phi$ modulation for photoproduced $\rho^{0}$, indicating a significant rescattering effect on all $\rho^{0}$. 

The origin of the negative $\cos 2\phi$ modulation for photoproduction can be understood from Fig.~\ref{fig:2}, and Fig.~\ref{fig:3} shows the simplified illustrations for the rescattering effect on hadronic $\rho^{0}$ and photoproduced $\rho^{0}$, respectively. The left panel of Fig.~\ref{fig:2} displays the spatial distribution of embedded $\rho^{0}$ mesons from photoproduction. A higher density of embedded $\rho^{0}$ is observed along the impact parameter direction ($x$ axis) compared to the reaction plane direction ($y$ axis). This anisotropy leads to stronger rescattering effects along the $x$ axis, resulting in a negative $\cos 2\phi$ modulation. The right panel of Fig.~\ref{fig:2} illustrates the $p_{T}$ distribution of embedded $\rho^{0}$ mesons from photoproduction. At this point, the case that satisfies these distributions is simply illustrated in the right panel of Fig.~\ref{fig:3}. It is indicated that due to the anisotropy of the nuclear overlapping regions and the $p_{T}$ distribution for the embedded $\rho^{0}$, the density of the hadron medium which the embedded $\rho^{0}$ passed through in the $x$ direction is greater than that in the $y$ direction, so the embedded $\rho^{0}$ in the $x$ direction will represent a stronger rescattering effect, and therefore the reconstructable $\rho^{0}$ that reflects the residual effect will lead to negative $\cos 2\phi$ modulation. In contrast, for hadronic $\rho^{0}$, as shown in the left panel of Fig.~\ref{fig:3}, it is generated in the nuclear overlapping regions. At this time, due to the anisotropy of the overlapping regions, the density of the medium along the $y$ direction is greater than that along the $x$ direction, so hadronic $\rho^{0}$ will produce a stronger rescattering effect in the $y$ direction. Therefore, the reconstructable $\rho^{0}$ that reflects the residual effect will lead to a positive $\cos 2\phi$ modulation.

Meanwhile, in the right panel of Fig.~\ref{fig:2}, it can also be indicated that the minimum in the distribution can be observed near the peak values at approximately 0.12 GeV/c and 0.20 GeV/c. This suggests that the number of reconstructable $\rho^{0}$ at these $p_{T}$ values is lower than in other lower $p_{T}$ regions. Additionally, the overall yield of coherent photoproduction is also reduced at these $p_{T}$ values. Consequently, the ratio of reconstructable to all $\rho^{0}$ increases at these $p_{T}$ peaks, leading to an enhancement to $\cos 2\phi$ modulation.

To explore the rescattering effect on embedding time, the $\cos 2\phi$ and $\cos 4\phi$ modulations as functions of embedding time for the case of random polarization are shown in Fig.~\ref{fig:4}. For reconstructable $\rho^{0}$ mesons from photoproduction, the $\cos 2\phi$ modulation remains largely unchanged for embedding times shorter than 1 fm/c. However, as the embedding time exceeds 1 fm/c, the modulation gradually decreases, approaching zero around 10 fm/c. Due to the short lifetime of the $\rho^{0}$ mesons ($\tau\approx1.3$ fm/c), even if the $\rho^{0}$ mesons are photoproduced at the beginning of the heavy-ion collisions, their decay daughters, $\pi^{\pm}$, can be produced approximately 1 fm/c later. Therefore, for embedding times shorter than 1 fm/c, the rescattering effects are expected to manifest after this time scale. The invariance of the modulation with respect to embedding time in this regime is therefore consistent with physical expectations.
 
Meanwhile, for embedding times beyond 1 fm/c, the decreasing density of the hadronic gas results in a diminishing rescattering effect, which explains the observed trend. 
 
The slight decrease observed in $\cos 2\phi$ modulation for embedding times shorter than 0.1 fm/c is mainly attributed to the statistic error resulting from photoproduced $\rho^{0}$ mesons colliding with protons and neutrons in the original colliding nucleus, a process referred to as a side reaction. This result further confirms that side reactions occurring at earlier embedding times do not influence the primary conclusions regarding the modulation behavior. Consistent with the observations in Fig.~\ref{fig:1}, the $\cos 4\phi$ modulation remains unaffected by rescattering effects.

\begin{figure}
    \centering
    \includegraphics[width=0.45\textwidth]{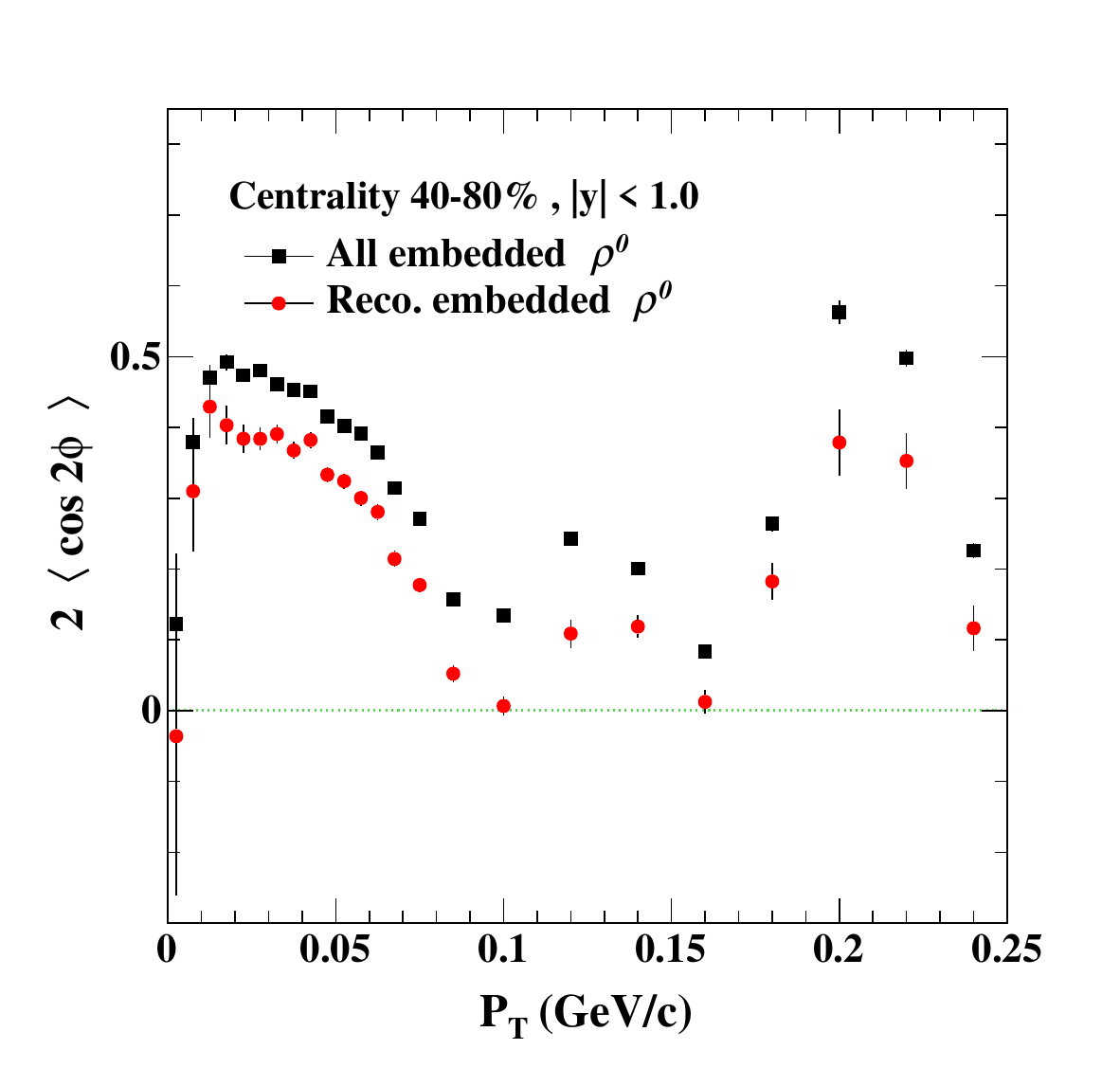}
    \caption{The $\cos2\phi$ modulation for the case of spin-interference from Ref.~\cite{ref:32} as a function of $p_{T}$ for reconstructable and all photoproduced $\rho^{0}$ at $40-80\%$ centrality in $\mathrm{Au}+\mathrm{Au}$ collisions at $\sqrt{s_{NN}}=200$ GeV at midrapidity with the fixed embeded time $0.01$ fm/c.}
    \label{fig:6}
\end{figure}

Similarly, we investigate the rescattering effect for photoproduced $\rho^{0}$ mesons in different centralities for the case of random polarization, as shown in Fig.~\ref{fig:5}. For the $\cos 2\phi$ modulation, it is observed that the rescattering effect is more significant in central collisions, while its significance diminishes in peripheral collisions. This trend is expected, as rescattering effect in the hadronic medium intensifies with increasing hadronic density towards central collision, leading to a more substantial modification of the $\cos 2\phi$ modulation in central collisions. In contrast, the $\cos 4\phi$ modulation shows little dependence on centrality and remains consistent with zero.

To accurately describe the physical conditions, it is necessary to incorporate the linear polarization of $\rho^{0}$ photoproduction into the angular distribution of the decay process within the UrQMD model. The degree of linear polarization for $\rho^{0}$ photoproduction can be calculated by the framework developed in our previous work~\cite{ref:32}. The implementation of linear polarization in the decay process is detailed in Section~\ref{method-3}. Fig.~\ref{fig:6} presents the $\cos 2\phi$ modulation as a function of $p_{T}$ at the fixed embedding time 0.01 fm/c for the case of linearly polarized $\rho^{0}$ photoproduction, calculated according to Ref.~\cite{ref:32} at $40-80\%$ centrality. As shown in the figure, the rescattering effect significantly modifies the modulation strength, although the overall shape of the distribution remains largely unchanged. This suppression of the $\cos 2\phi$ modulation reflects the effect of hadronic interactions in the medium, which influences the decay of photoproduced $\rho^{0}$ mesons. The fact that the overall distribution shape is preserved suggests that the underlying dynamics of photoproduction remains intact, with the rescattering effect only modifying the relative intensity of the angular modulations.

\section{Conclusion}
\label{sec:4}

In conclusion, we have investigated the rescattering effect on spin-interference observables for photoproduced $\rho^{0}$ mesons in $\mathrm{Au} + \mathrm{Au}$ collisions at $\sqrt{s_{NN}} = 200$ GeV using the modified UrQMD model. Our results demonstrate that the $\cos 2\phi$ modulation exhibits significant negative values under random polarization conditions, which are substantially altered when spin-interference is taken into account. The rescattering effect on the $\cos 2\phi$ modulation weakens with increasing embedding time and in more peripheral collisions, while the $\cos 4\phi$ modulation remains largely unaffected by both embedding time and centrality. These findings indicate that the rescattering effect significantly reduces the observable spin-interference signal from photoproduction. In the experiment, we can distinguish the $\rho^{0}$ from hadronization and photoproduction by setting the $p_{T}$ cut of $\pi^{\pm}$ pairs. The coherent photoproduced $\rho^{0}$ are basically $p_{T}<0.3$ GeV/c. Within this range, the yield from hadronization is more lower. Therefore, we can measure the $\cos 2\phi$ modulation from photoproduction in peripheral collisions in the experiment. In this sense, it is essential to account for the difference between experimental measurements and theoretical calculations for spin-interference from $\rho^{0}$ photoproduction in heavy-ion collisions.

\section{Acknowledgments}

The authors thank Dr.~Daniel Brandenburg and Dr.~Leszek Kosarzewski for helpful discussions. This work is supported in part by the National Key Research and Development Program of China under Contract No. 2022YFA1604900 and the National Natural Science Foundation of China (NSFC) under Contract No. 12175223 and 12005220. W. Zha is supported by Anhui Provincial Natural Science Foundation No. 2208085J23 and Youth Innovation Promotion Association of Chinese Academy of Science.

\bibliographystyle{unsrt}
\bibliography{reference.bib}

\end{document}